\IfFileExists{IEEEtran.cls}{%
	\documentclass[conference]{IEEEtran}%
	\newcommand{\HASIEEE}{}%
}{%
	\documentclass[10pt,twocolumn]{article}%
	\usepackage[margin=0.75in]{geometry}%
}
\ifdefined\HASIEEE
\IEEEoverridecommandlockouts
\else
\newcommand{\IEEEauthorblockN}[1]{#1}
\newcommand{\IEEEauthorblockA}[1]{\\ \small #1}
\newenvironment{IEEEkeywords}{\par\smallskip\noindent\textbf{Index Terms---}}{\par}
\fi

\usepackage{cite}
\usepackage{amsmath,amssymb}
\usepackage{graphicx}
\graphicspath{{./figures/}}
\usepackage{booktabs}
\usepackage{makecell}
\usepackage{multirow}
\usepackage[hyphens]{url}
\usepackage[hidelinks]{hyperref}
\usepackage{xcolor}
\begin{document}
	
	\title{Separating Voice from Age in COPD Screening}
	
	\author{\IEEEauthorblockN{George P. Kafentzis, Nikoletta Arvaniti}
		\IEEEauthorblockA{Department of Computer Science, University of Crete}}
	
	\maketitle
	
	\begin{abstract}
		Voice has been proposed as a low-cost screening signal for chronic obstructive
		pulmonary disease (COPD). COPD is strongly age-associated and voice changes with
		age, thus such results admit a trivial alternative explanation. We re-evaluate a
		public sustained-phonation corpus ($1246$ recordings, $68$ participants) under a
		strictly participant-level protocol. We	therefore evaluate on repeatedly drawn age-matched cohorts and report the discrimination achieved by the confounders themselves on those same cohorts. Where raw (unmodelled) age ($0.510$ $[0.469, 0.551]$) and raw gender ($0.479$) are both measured at chance, acoustic models excluding age retain ROC-AUC $0.717$ $[0.552, 0.859]$ and average precision $0.747$ $[0.581, 0.892]$ against a one-to-one baseline of $0.5$, whereas models containing age fall to $0.531$--$0.679$. The separation is reproduced 
		by two further learners with fixed hyperparameters. Two findings have broader methodological implications: models trained with age transfer less effectively to an age-balanced target cohort than otherwise identical models trained without age, and fourteen classical voice-quality and perturbation measures achieve comparable discrimination to a $55$-dimensional combined representation. We conclude that a non-age acoustic signal is present, that
		confounding by recording conditions cannot be excluded from the released
		features, and that the evaluation protocol in standard use cannot distinguish
		these possibilities.
	\end{abstract}
	
	
	\section{Introduction}
	
	Chronic obstructive pulmonary disease (COPD) is a progressive, long-term lung disease that blocks airflow and makes breathing difficult~\cite{who2026copd}. It primarily encompasses two main conditions: emphysema (damage to the tiny air sacs in the lungs) and chronic bronchitis (ongoing inflammation and mucus buildup in the airways). COPD affects several hundred million people worldwide and is among the leading causes of death~\cite{boers2023prevalence,gold2023report}. Early-stage COPD is frequently asymptomatic, thus a large fraction of cases remain undiagnosed until irreversible airway damage has occurred. Spirometry is the diagnostic reference standard, but it requires clinic attendance, trained personnel, and patient effort, which limits its use for preliminary	screening at population scale.
	
	Voice has therefore attracted attention as a candidate low-cost, non-invasive
	screening signal~\cite{vanbemmel2026tacticas}. The rationale is physiologically motivated: COPD reduces	expiratory flow and vital capacity, and phonation depends directly on subglottal pressure and airflow. Reduced breath support, altered vocal-fold
	vibration, and increased spectral noise are consequently plausible acoustic
	correlates of the disease, and sustained phonation of a vowel is the simplest
	task in which such effects can be observed. A growing literature reports that
	machine-learning classifiers can separate COPD patients from healthy controls
	on the basis of such recordings, with accuracies commonly in the $75$--$95\,\%$ range~\cite{idrisoglu2024copdvd,sankeyolsen2025danish,ankishan2025voice,obase2026biometric}.
	
	This paper is concerned with a question that is logically prior to how well	such classifiers perform: \emph{what are they measuring?} COPD is strongly age-associated. Voice also changes substantially with age, through vocal-fold atrophy, altered mucosal wave, reduced respiratory support, and changes in fundamental frequency and its variability~\cite{VoiceAge, VoiceAge2}. Any cohort in which cases are older than controls therefore admits a trivial alternative explanation for	classification performance: the model may be estimating age, or any acoustic correlate of age, rather than disease. This concern is neither novel nor subtle, and it is routinely acknowledged in the speech-health literature~\cite{ramanarayanan2022speech}. What is less common is a design that	actually tests it.
	
	The difficulty is that the usual approach, e.g. reporting a model trained with the
	age variable removed, is insufficient. If acoustic features are themselves
	correlated with age, removing this single number does not remove the information it carries. The reason is that excluding a potentially confounding variable from the predictor set does not necessarily eliminate its influence, since information about that variable may remain redundantly encoded in correlated predictors, allowing machine-learning models to exploit proxy representations of the omitted variable~\cite{cornacchia2023auditing, barocas2023fairness, datta2017use}. In our case, a sufficiently flexible learner	may be able to reconstruct an age-driven decision boundary from age-correlated acoustics. Establishing that a voice model carries disease information beyond age requires either (i) demonstrating that the acoustic features are not age-proxies, or (ii) evaluating on cohorts in which age showed negligible monotone discrimination and small mean imbalance. In this work, we examine the former and explicitly implement the latter.
	
	We work with a recent, publicly described Swedish sustained-phonation dataset
	of $1246$ recordings from $68$ participants, collected and released with the
	COPDVD study~\cite{idrisoglu2024copdvd}. Each recording is represented by $107$ predictors: age and gender, four self-reported throat-condition variables (cold, pain, phlegm, and other), $49$ static acoustic descriptors, including jitter, shimmer, harmonics-to-noise ratio, $F_0$ statistics, phonation duration, formant statistics, and Mel Frequency Cepstral Coefficients (MFCC) statistics, and $52$ summary features derived from first- and second-order MFCC temporal derivatives. Table~\ref{tab:features} summarizes the features.
	
	\begin{table}[t]
		\centering
		\caption{Features used for COPD classification. Refer to~\cite{idrisoglu2024copdvd} for further details.}
		\label{tab:features}
		\footnotesize
		\setlength{\tabcolsep}{6pt}
		\renewcommand{\arraystretch}{1.05}
		\begin{tabular}{llc}
			\toprule
			\textbf{Feature group} & \textbf{Feature} & \textbf{No.} \\
			\midrule
			
			\multirow{2}{*}{Demographic}
			& Age    & 1 \\
			& Gender & 1 \\
			\midrule
			
			\multirow{4}{*}{Health}
			& Cold (Cold/Flu)       & 1 \\
			& Pain (Sore throat)    & 1 \\
			& Slimy (Mucus in throat) & 1 \\
			& Other                 & 1 \\
			\midrule
			
			\multirow{9}{*}{\makecell[l]{Baseline\\ Acoustic\\Features}}
			& Duration                         & 1 \\
			& Mean $F_0$                       & 1 \\
			& Std. $F_0$                       & 1 \\
			& HNR                              & 1 \\
			& Local, Absolute, RAP, PPQ5, DDP jitter
			& 5 \\
			& Local, local-dB shimmer           & 2 \\
			& APQ3, APQ5, APQ11, DDA shimmer   & 4 \\
			& $F_1$--$F_4$ means                & 4 \\
			& $F_1$--$F_4$ medians              & 4 \\
			\midrule
			
			\multirow{6}{*}{\makecell[l]{Mel-Frequency\\Cepstral\\Coefficients\\(MFCCs)}}
			& MFCC$_{1:13}$ mean                & 13 \\
			& MFCC$_{1:13}$ std.                & 13 \\
			& $\Delta$MFCC$_{1:13}$ mean        & 13 \\
			& $\Delta$MFCC$_{1:13}$ std.        & 13 \\
			& $\Delta^2$MFCC$_{1:13}$ mean      & 13 \\
			& $\Delta^2$MFCC$_{1:13}$ std.      & 13 \\
			\midrule
			
			\multicolumn{2}{r}{\textbf{Total}} & \textbf{107} \\
			\bottomrule
		\end{tabular}
	\end{table}
	
	This paper contributes the following:
	
	\begin{enumerate}
		\item \textbf{A dataset audit:} The cohort is described in the source publication as matched on age and gender within a five-year range. At participant level this matching does not hold: the standardised mean difference in age between cases and controls is $0.73$, eight of $37$ controls have no case within five years, and ten fall outside the case age range entirely. The discrepancy arises because descriptive statistics weighted by \emph{recording} rather than by \emph{participant} invert the apparent direction of the age difference. 
		
		\item \textbf{A demonstration that pooled evaluation is uninformative under this confounding:} With age imbalance present, demographic-only, acoustic, and combined models are statistically indistinguishable, and a hand-specified non-monotone function of age alone matches the best model. No incremental comparison in the conventional design resolves a nonzero effect. This is a property of the evaluation protocol, not of the data.
		
		\item \textbf{An age-matched evaluation protocol with verified controls:}
		We evaluate several configurations on repeatedly drawn one-to-one age-matched
		cohorts and report, on each cohort, the discrimination achieved by \textit{raw} (not fitted) age	and raw gender. We show that a \emph{fitted} age-only model is not a valid positive control for this purpose, because its out-of-sample predictions
		carry a cross-fitting-induced anti-correlation artefact, and we quantify that
		artefact.
		
		\item \textbf{Evidence for a residual acoustic signal:} On \textit{matched cohorts}
		where raw age and raw gender both discriminate at chance, acoustic models
		excluding age retain an Area under the Receiver Operating Curve (ROC-AUC) $0.665$--$0.717$ and average precision $0.734$--$0.755$, while age-containing models fall to $0.531$--$0.679$. The dissociation replicates in direction under two further learners. 
		
		\item \textbf{A methodological finding of wider scope:} Supplying a confounded covariate during training does not leave the acoustic representation intact: models given age preferentially exploit it, under-learn the acoustic structure, and the deficit persists once the confound is neutralized. This argues for excluding confounded demographics	at training time rather than reporting a post-hoc ablation.
		
		\item \textbf{Feature compression:} Fourteen classical perturbation	measures achieve comparable discrimination to a $55$-dimensional combined representation. Specifically, MFCC temporal derivatives, symptom questionnaire items, and gender each contribute nothing detectable. We also document that low-dimensional configurations are unstable to hyperparameter selection noise, and adopt a one-standard-error rule that removes it.
	\end{enumerate}
	
	We should emphasize what this paper does \emph{not} claim. It does not claim that voice-based COPD screening is clinically viable, at least at this level of dataset sizes. The matched cohorts contain roughly two dozen pairs, drawn from a single site, a single language, and a single recording protocol. The claim is narrower and
	methodological: a non-age acoustic signal is present and measurable in these data, and the evaluation protocol in standard use cannot see it.
	
	\section{Related Work}
	
	We group prior work by the question it asks, because the confounding structure
	differs fundamentally between designs. Section~\ref{sec:rw-cross} covers
	cross-sectional case--control detection, where between-subject confounding is
	the dominant threat. Section~\ref{sec:rw-long} covers within-patient longitudinal monitoring, where each patient serves as their own control and the threat largely disappears. Section~\ref{sec:rw-confound} covers work that addresses speaker identity and confounding explicitly. Finally, Section~\ref{sec:rw-synth} relates current and previous works.
	
	\subsection{Cross-Sectional COPD Detection from Sustained Phonation and Read Speech}
	\label{sec:rw-cross}
	
	Idrisoglu et al.~\cite{idrisoglu2024copdvd} collected the COPDVD corpus---$1246$ recordings of the sustained vowel /a/ from $68$ Swedish participants ($30$ COPD, $38$ controls), captured on participants' own smartphones over six months. They compared Random Forest, SVM, and CatBoost using nested cross-validation with grid search. CatBoost performed best, reaching $78\,\%$ test accuracy, ROC-AUC $0.82$, and average precision $0.76$, with SHAP analysis ranking age, gender, and the means of baseline acoustic and MFCC features most influential. This is, to our knowledge, the most carefully constructed public resource of its kind. The authors partition at participant level, apply nested cross-validation explicitly to limit selection bias, compare
	several model families rather than reporting a single tuned result, publish
	feature-importance analyses, and provide the dataset upon request. 
	Age and gender are included as model inputs and reported among the most important features, but their contribution is not separated from the acoustic contribution. Thus, the	model performance cannot be attributed to voice alone. 
	
	Idrisoglu et al.~\cite{idrisoglu2025segmentation} follow up by asking how segmentation of the sustained vowel affects classification, comparing full-sequence, segment-wise, and group-wise dataset constructions across nested cross-validation configurations from $2\times2$ to $5\times5$. The question is well posed and practically important: if
	different portions of a sustained vowel carry different amounts of information, both feature design and recording protocol should reflect it. 
	On the other hand, segment-wise dataset construction creates a direct route to optimistic bias, because segments originating from the same recording (same speaker, session, and device) may be distributed across training and evaluation folds. The reported gap between validation ($97.8\,\%$) and test ($84.6\,\%$) accuracy is again wide. 
	
	Sankey-Olsen et al.~\cite{sankeyolsen2025danish} present a Danish corpus from $96$ participants and evaluate MFCC features alongside learned x-vector embeddings~\cite{snyder2018xvectors}. In this work, a new corpus in an under-served language is introduced, combining hand-crafted and learned representations, with recordings collected both in person and remotely, suggesting a more realistic picture of deployment conditions. However, x-vectors are trained for speaker discrimination and therefore encode speaker identity by construction: demographic attributes such as age and sex may consequently remain recoverable from the representation~\cite{kwasny2021gender}. Using them for a between-subject health classification makes the identity/pathology confound maximal rather than minimal, and the study does not attempt to
	separate the two.
	
	Anki\c{s}han et al.~\cite{ankishan2025voice} classify healthy controls, lung cancer, and COPD from Turkish read speech ($n=128$: $42$ controls, $50$ lung cancer, $36$ COPD) using combined spectral, cepstral, and learned-embedding features with an XGBoost classifier and SHAP analysis. The three-way design is more informative than binary case--control, because it forces the model to distinguish two diseases from each other and not merely sick from healthy. The authors use a stratified \emph{subject-level} $80/20$ split explicitly to prevent subject leakage, and combine it with $5$-fold cross-validation for hyperparameter selection. However, a single held-out split of $128$ participants yields a test set too small for stable estimates, and no interval estimates are reported. The lung-cancer and COPD groups differ from controls in age and smoking history, and these are not controlled, so the same confounding concern applies here as elsewhere.
	
	Obase et al.~\cite{obase2026biometric} recorded voice and cough from $55$ Japanese participants ($26$ COPD, $29$ non-COPD) with pulmonary function testing, comparing smartphone and dedicated recorders. This is the most methodologically cautious study we have found. The authors use leave-one-out cross-validation explicitly to avoid artificially inflating the effective sample size, they report ROC-AUC with sensitivity and specificity rather than accuracy alone, and they state plainly that overall sensitivity and specificity fall below $80\,\%$ and that diagnostic performance is therefore limited. The comparison of recording devices addresses a real deployment question, and augmenting acoustics with	COPD Assessment Test scores tests whether voice adds to what a questionnaire already provides. It should be noted that the headline ROC-AUC of $0.828$ is achieved in a male-restricted subgroup, which reduces the effective sample further. Also, age differences between groups are not explicitly controlled.
	
	Leyva-Bravo et al.~\cite{leyvabravo2024vocal} apply a Transformer-based classifier to $30$ vocal recordings from vulnerable communities in Guerrero, Mexico, including participants exposed to biomass smoke. The population is under-studied, and biomass-smoke exposure is a major and under-represented COPD aetiology. Including an exposed-but-not-diagnosed group is conceptually valuable, since it probes whether the model responds to disease or to exposure.
	
	\subsection{Within-Patient Longitudinal Monitoring}
	\label{sec:rw-long}
	
	A distinct and, in our view, better-posed line of work asks not whether a speaker has COPD but whether \emph{this} patient's condition has changed. Because each patient serves as their own control, age, sex, anatomy, and habitual voice quality are held fixed by construction, and the between-subject confounding that dominates Section~\ref{sec:rw-cross} largely vanishes.
	
	van Bemmel et al.~\cite{vanbemmel2021features} compare Dutch COPD patients in exacerbation ($n=11$), in stable condition ($n=9$), and healthy controls ($n=29$) using read speech, ranking Praat and eGeMAPS features by recursive feature elimination and classifying with SVM and LDA. The authors use leave-one-speaker-out cross-validation and state explicitly that this is to reduce identity confounding. They report Matthews correlation coefficient (MCC) rather than accuracy because the classes are severely imbalanced, and they match reference recordings to COPD recordings on duration. The features that pass selection (vocalization-to-inhalation ratio, average phonation time, pause frequency) are physiologically interpretable and directly related to breath support. It should be noted, however, that the sample is very small.
	
	Nallanthighal et al.~\cite{nallanthighal2022exacerbation} detect COPD exacerbation from speech in $40$ patients, comparing openSMILE acoustic features with deep speech-breathing models that estimate breathing signals from audio, reaching $75.12\,\%$ accuracy and $0.85$ sensitivity. The speech-breathing model is a novel methodological contribution: rather than treating acoustics as an opaque feature vector, it estimates a physiologically meaningful latent variable, the breathing rate, and shows it differs between exacerbated and stable states, leading to a credible biomarker. However, since all patients have COPD, this work contributes to \emph{monitoring} rather than \emph{screening}.
	
	van Bemmel et al.~\cite{vanbemmel2026tacticas} report the TACTICAS study: $73$ participants ($35$ asthma, $38$ COPD, mean age $62$) providing daily home voice recordings, capturing $38$ exacerbations from $35$ participants. This work illustrates the conditions under which such a system would actually be deployed. Adherence is measured and reported, which most speech-health work omits and which determines whether a monitoring system is viable at all.
	As a within-patient design, it cannot address screening in undiagnosed populations. 
	
	Mayr et al.~\cite{mayr2024clinical} compare speech during and after exacerbation in $50$ COPD patients, reporting $84\,\%$ accuracy and finding speech analysis superior to BORG and CAT scores alone. The comparison against established clinical instruments is very interesting, since the question is not whether speech carries signal but whether it adds to what clinicians already measure cheaply. The paired within-patient design is strong, and the GOLD-stage distribution is reported. A drawback is that the study is single-center, as almost all similar works.
	
	Bhalla et al.~\cite{bhalla2025wearable} analyse real-life speech captured by smartwatch from $18$ COPD patients over an average of roughly $200$ days, using linear mixed-effects models to relate phonation and prosodic features to daily lung condition. The authors explicitly model demographic covariates as confounders and examine physiological moderators such as heart-rate variability. Again, this study emphasizes monitoring rather than screening.
	
	Farr\'{u}s et al.~\cite{farrus2021support} present a speech-based COPD supervision system in Spanish, with home recordings at rest and after exercise. Classification uses $10$-fold cross-validation over \emph{instances}, chosen explicitly to maximise the number of training instances. Multiple recordings per participant are distributed across folds, the reported performance includes an unknown component of speaker recognition.
	
	Ali Khan et al.~\cite{alikhan2021cat} predict deviation in COPD Assessment Test score from $418$ voice sessions collected from $9$ participants, reporting $R^2 = 0.972$, $r = 0.998$, and $94\,\%$ accuracy under $k$-fold cross-validation. This work targets a continuous, clinically meaningful outcome rather than a binary label. However, with $418$ sessions from $9$ participants and folds that are not speaker-disjoint, a correlation of $0.998$ should be interpreted cautiously.
	
	\subsection{Speaker Identity and Confound Control}
	\label{sec:rw-confound}
	
	Yan et al.~\cite{yan2026disentangled} address the identity problem directly, training a speaker-disentangled representation on TACTICAS data with adversarial gradient reversal to suppress speaker information while optimising respiratory-status classification. Disentanglement raises stable-versus-exacerbated AUC from $0.897$ to $0.910$, and asthma-versus-COPD exacerbation AUC from $0.674$ to $0.793$, while reducing a measured identity-leakage ratio.	The authors recognise that speaker identity is a confounder rather than a nuisance, they measure leakage explicitly rather than assuming it away, they use speaker-disjoint train/validation/test partitions, and they demonstrate the practically significant result that suppressing identity \emph{improves} rather than degrades diagnostic accuracy. However, the approach suppresses speaker identity \emph{in the representation}, which is the right target for a longitudinal setting where the same speakers recur. It does not address	\emph{group-level demographic imbalance} in a cross-sectional cohort. If cases are systematically older than controls, a representation from which individual identity has been removed may still encode age, and age remains predictive of the label. 
	
	Since this work is closer to ours, our contribution is complementary: rather than modifying the representation, we modify the evaluation, constructing cohorts on which the confounder is verifiably uninformative and reporting the confounder's own discrimination as a control.

	\subsection{Synthesis and the Basis of the Present Work}
	\label{sec:rw-synth}
	
	Two observations follow from the summary of related work. First, \textit{the field is split by design in a way that determines its exposure to confounding}. The longitudinal studies
	\cite{vanbemmel2021features,nallanthighal2022exacerbation,vanbemmel2026tacticas,mayr2024clinical,bhalla2025wearable}
	are largely immune to between-subject confounding, and correspondingly report the most physiologically grounded findings such as breathing rate, phonation time, vocalization-to-inhalation ratio. The cross-sectional screening studies
	\cite{idrisoglu2024copdvd,idrisoglu2025segmentation,sankeyolsen2025danish,ankishan2025voice,obase2026biometric,leyvabravo2024vocal}
	are fully exposed to it, and report higher headline numbers. The two groups	are frequently cited together as evidence for the same proposition, which they are not: performance on a within-patient comparison provides no support for screening in undiagnosed populations. Second, \textit{where confounding is addressed, it is addressed at the level of the individual rather than the group}. Speaker-disjoint partitioning \cite{vanbemmel2021features,ankishan2025voice,yan2026disentangled} prevents a model from recognizing a specific speaker seen in training. Adversarial disentanglement \cite{yan2026disentangled} goes further and removes identity from the representation. Neither addresses the case in which the case and control \emph{groups} differ demographically. A perfectly speaker-disjoint, perfectly identity-disentangled model evaluated on a cohort where patients are eight years older than controls will still exploit age, and will still report strong performance.
	
	We are aware of no prior work in COPD voice analysis that (i) verifies the demographic balance of the cohort it evaluates on, (ii) reports the	discrimination achieved by the confounder itself as a control condition, or (iii) distinguishes what a model has learned from what a confound would predict. Where matching is claimed, it is asserted from cohort-level summary statistics rather than demonstrated on the analyzed partition. We show in Section~\ref{sec:audit} that summary statistics can conceal, and even invert,
	the imbalance they are meant to characterize.
	
	The present work therefore proceeds as follows. We adopt a strictly	participant-level protocol throughout: grouped outer and inner cross-validation, inverse recording-count weighting so that participants contribute equally regardless of how many recordings they provided, subject-level aggregation to a single score per participant, and bootstrap
	resampling over participants rather than recordings. Within that protocol we first show that conventional pooled evaluation cannot answer the question of interest on this cohort, then introduce an age-matched evaluation with verified controls that can, and finally replicate the result under an independent learner.

	\section{Dataset Audit}
	\label{sec:audit}
	
	Before any model is fitted, we characterize the cohort at the level at which
	classification is performed. This section reports what we found, and why the
	conventional description of the corpus does not survive that examination.
	
	\begin{figure*}[htb!]
		\centering
		\includegraphics[width=\textwidth]{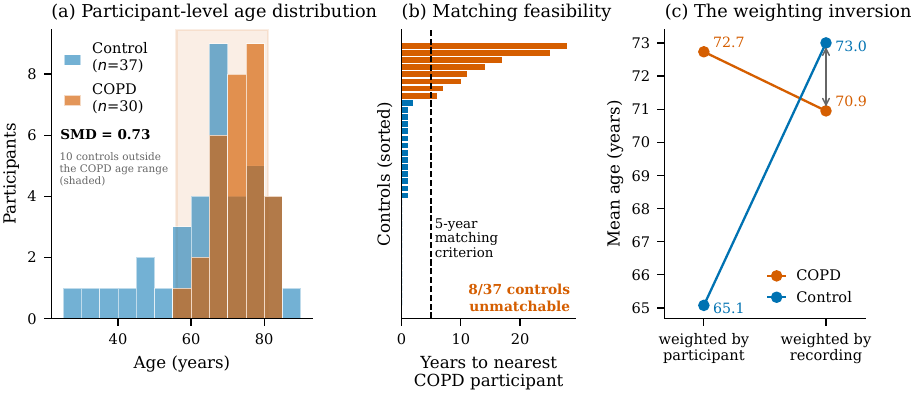}
		\caption{Age structure of the cohort. (a)~Participant-level age distributions;
			the shaded band marks the COPD age range, outside which ten controls fall.
			(b)~For each control, the age distance to the nearest COPD participant; eight
			of $37$ exceed the five-year matching criterion stated in the source
			publication and are shown in orange. (c)~Group mean age computed over
			participants and over recordings, on the full $68$-participant corpus. The sign
			of the group difference reverses between the two weightings.}
		\label{fig:age}
	\end{figure*}
	
	\subsection{Corpus and Analyzed Cohort}
	
	The COPDVD corpus comprises $1246$ recordings of the sustained vowel /a/ from $68$
	participants ($30$ COPD, $38$ controls), captured on participants' own smartphones over a six-month period~\cite{idrisoglu2024copdvd}. Each recording is represented by $107$ predictors: age and gender, four self-reported throat-condition items recorded immediately before each session (cold, pain, phlegm, other), $49$ static acoustic descriptors comprising jitter ($5$), shimmer ($6$), harmonics-to-noise ratio, mean and standard deviation of $F_0$, phonation duration, four formant means and medians, and the means and standard deviations of $13$ MFCCs. Also, $52$ first- and second-order MFCC temporal derivative summaries are included.
	
	Recording counts per participant are severely unbalanced. One control participant contributes $306$ recordings, $24.6\,\%$ of the entire corpus, against a median of $8$ and a next-largest count of $69$. Retaining this participant makes grouped fold construction degenerate: any partition either places a quarter of the data in a single test fold or leaves that fold without representation from the largest contributor. We therefore exclude this participant, giving an analysed cohort of $67$ participants and $940$ recordings ($30$ COPD, $37$ controls; prevalence $0.448$), with the largest
	single contribution reduced to $7.3\,\%$. We report this exclusion explicitly because, as shown below, the excluded participant is material to how the	cohort's age structure appears.
	
	\subsection{The Age Imbalance}
	
	The source publication states that participants were matched on age and gender within a five-year range, and reports group mean ages of $72.04$ and $72.13$	years for COPD and controls respectively---a difference of less than one month~\cite{idrisoglu2024copdvd}.
	
	At participant level, the analyzed cohort does not exhibit this balance. COPD participants have mean age $72.73$ years (SD $6.34$, range $56$--$81$) while controls have mean age $64.68$ years (SD $14.37$, range $28$--$88$). The	standardized mean difference is defined as
	\begin{equation}
		\mathrm{SMD} = \sqrt{2}\frac{\bar{x}_{COPD} - \bar{x}_{HC}}
		{\sqrt{s_{COPD}^{2} + s_{HC}^{2}}}
		\label{eq:smd}
	\end{equation}
	where $\bar{x}_{COPD}$ and $\bar{x}_{HC}$ denote the sample means of age variable $x$
	in the COPD and healthy control (HC) groups, respectively, while $s_{COPD}$ and $s_{HC}$ denote the corresponding sample standard deviations. For our cohort, SMD equals $0.726$. A value above $0.1$ is conventionally taken to indicate meaningful imbalance in matched-cohort work~\cite{austin2011caliper}. Thus, $0.726$ is a large imbalance by any standard. 
	
	The imbalance is structural rather than a matter of means. Fig.~\ref{fig:age}(a)
	shows that the two age distributions differ in dispersion as much as in	location: COPD participants span a narrow $25$-year window, while controls span $60$ years, including six participants under $50$ for whom no COPD counterpart of any age exists. Ten controls fall outside the COPD age range entirely. Fig.~\ref{fig:age}(b) makes the consequence for matching explicit: for each	control we plot the age distance to the nearest COPD participant. Eight of the $37$ controls have no COPD participant within five years, and are therefore unmatchable under the stated criterion.
	
	The unit over which descriptive statistics are computed provides a plausible explanation for the discrepancy between the reported and observed balance. Fig.~\ref{fig:age}(c)
	contrasts the two weightings on the full $68$-participant corpus. Weighted by
	participant, COPD participants are $7.7$ years older than controls
	($72.73$ versus $65.08$). Weighted by recording, controls are $2.1$ years
	\emph{older} than COPD participants ($73.00$ versus $70.95$). The sign of the
	group difference reverses. Two properties of the corpus produce this. First, the $80$-year-old control contributing $24.6\,\%$ of recordings shifts the recording-weighted control mean upward by more than four years on its own. Second, recording count	correlates with age in opposite directions within the two groups
	($\rho = +0.24$ among controls, $\rho = -0.27$ among COPD participants), so
	older controls and younger patients are both over-represented at recording
	level. The recording-weighted figures are close to those reported in the source
	publication, which is consistent with the reported cohort description having
	been computed over recordings rather than over participants.
	
	This is not a criticism specific to one corpus. Any repeated-measures speech
	dataset with unequal per-participant contributions admits the same failure, and
	the direction of the distortion is not predictable in advance. It follows that
	cohort balance must be reported at the unit of analysis used by the classifier.
	Where a model makes one prediction per participant, participant-level balance
	is the quantity that matters, and recording-level summaries can conceal---and
	here invert---the imbalance they are intended to characterize.
	
	\subsection{The Age--Disease Relationship}
	
\begin{figure}[t]
		\centering
		\includegraphics[width=\columnwidth]{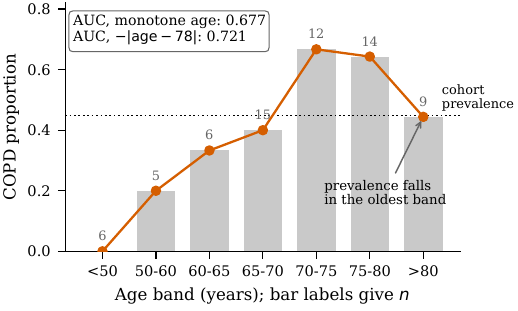}
		\caption{COPD proportion by age band ($n$ above each bar). Prevalence peaks in
			the $70$--$75$ band and declines thereafter, because the COPD group is bounded
			above at $81$ years. A non-monotone score in age alone attains ROC-AUC $0.721$,
			against $0.677$ for a monotone score.}
		\label{fig:nonmono}
	\end{figure}

	Fig.~\ref{fig:nonmono} shows COPD proportion by age band. Prevalence rises from
	zero below $50$ to $0.67$ in the $70$--$75$ age band, then \emph{falls} to $0.44$
	above age $80$. The reversal is a recruitment artefact: the COPD group is upper-bounded
	at $81$ years, so the three oldest participants in the cohort are
	necessarily controls.
	
	This structure has a direct consequence for how much of the classification task
	age alone can account for. All COPD-control pairs are $30 \times 37 = 1110$ and ROC-AUC is a rank statistic. Thus, a monotone score in age reaches ROC-AUC $0.677$ (as it wins $752$ out of $1110$ pairs), but a trivially specified non-monotone score, such as $-\lvert \mathrm{age} - 78 \rvert$, reaches $0.721$ (it wins $800$ out of $1110$ pairs). Any learner able to represent a non-monotone function of age---which includes every tree ensemble---therefore has access to approximately $0.72$ ROC-AUC from the age variable alone. We return to this in	Section~\ref{sec:results}, where it establishes that pooled evaluation on this cohort cannot separate acoustic from demographic information.
	
	\subsection{Acoustic Features as Age Proxies}
	
	It is interesting to investigate whether removing the numerical value of age does or does not remove age information. That explicitly depends on whether the acoustic features are themselves age-dependent. We therefore quantify the association directly. For each acoustic descriptor we compute the	participant-level Spearman correlation with age, alongside its univariate discrimination of COPD.
	
\begin{figure}[t]
		\centering
		\includegraphics[width=\columnwidth]{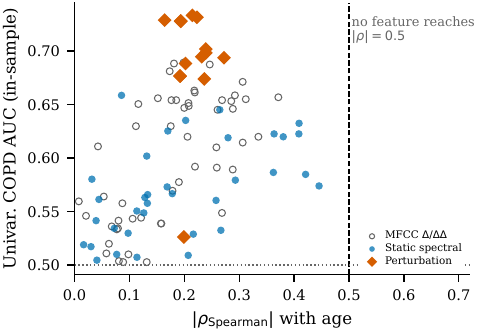}
		\caption{Association with age against univariate COPD discrimination, for every
			acoustic descriptor at participant level. The strongest discriminators
			(perturbation measures, orange diamonds) are among the least age-correlated;
			the most age-correlated features (MFCC standard deviations) discriminate
			weakly. Univariate AUCs are in-sample and serve to order features, not to
			estimate performance.}
		\label{fig:proxy}
	\end{figure}

	Fig.~\ref{fig:proxy} shows the result of this experiment. Across the $49$ static descriptors the median $\lvert\rho\rvert$ with age is $0.19$ and the maximum is $0.45$. Moreover, seven descriptors exceed $0.3$ and none reaches $0.5$. The representation is therefore only weakly to moderately age-dependent. The most age-correlated features are the MFCC standard-deviation family ($\lvert\rho\rvert$ between $0.36$ and $0.45$ --- blue/filled dots on the right part of the figure), which are comparatively weak discriminators. The strongest univariate discriminators are the perturbation measures---\texttt{localJitter} ($0.733$), \texttt{ppq5Jitter} ($0.732$), \texttt{sd\_F0\_list} ($0.729$), \texttt{ddpJitter} and \texttt{rapJitter} (both $0.728$)---which belong to the baseline acoustic features of Table~\ref{tab:features}. These are represented by the group of orange rhombuses at the top of Fig.~\ref{fig:proxy}, and these are among the \emph{least} age-correlated ($\lvert\rho\rvert$ between $0.16$ and $0.24$). Hence, the strongest marginal COPD discriminators and the strongest marginal age correlates occupy largely different parts of the feature space.
	It should be clearly stressed that the univariate AUCs are computed in-sample and are
	optimistically biased: they are reported to establish the relative ordering of
	features, not their absolute performance. Also, weak marginal correlation with
	age does not preclude a multivariate combination of features from recovering
	age more effectively than any single feature does. This analysis therefore
	motivates the matched evaluation of Section~\ref{sec:matched} rather than
	substituting for it.
	\subsection{Other Candidate Confounders}
	
	Gender is not a confounder in this cohort: $46.7\,\%$ of COPD participants and
	$48.6\,\%$ of controls are male, and gender as a raw (unmodeled) score achieves ROC-AUC
	$0.490$, marginally below chance. Recording count is likewise uninformative
	(ROC-AUC $0.545$), indicating no systematic difference in study adherence
	between groups.	Phonation duration is the one acoustic variable requiring comment. Mean
	duration is $15.26$\,s for COPD participants and $16.71$\,s for controls, giving
	ROC-AUC $0.602$. Under a sustain-as-long-as-possible instruction this is
	maximum phonation time, a recognized clinical correlate of reduced vital
	capacity~\cite{feltrin2026maximum}, and therefore legitimate signal: under a fixed-duration instruction the same difference would be a protocol artefact. We cannot resolve this from the released features, and thus we report an acoustics-only configuration with duration removed throughout, so that no conclusion rests on it.

	\section{Methods}
	\label{sec:methods}
	
	The design goal throughout is that every quantity we report should be
	interpretable at the level of the participant, since that is the unit at which
	a screening decision would be made. This constrains fold construction, sample
	weighting, score aggregation, and resampling, each of which is described below.
	Section~\ref{sec:matched} then describes the matched-cohort evaluation, which is
	the component that distinguishes this work from prior evaluations of the same
	corpus.
	
	\subsection{Feature Configurations}
	\label{sec:configs}
	
	We evaluate twelve nested feature configurations, listed in
	Table~\ref{tab:configs}. They are constructed to isolate the contribution of
	each block rather than to maximise performance. Four contrasts are of primary
	interest: (i) \texttt{A} against the acoustic configurations bounds what age alone
	can achieve, (ii) the pairs $\texttt{DAC+}\Delta$ against \texttt{DAC}, and
	$\texttt{ALL+}\Delta$ against \texttt{ALL}, isolate the MFCC temporal
	derivatives,  (iii) the pairs \texttt{DAC} against \texttt{DAC-A}, and \texttt{ALL}
	against \texttt{ALL-A}, isolate the explicit age variable, and (iv) \texttt{DAC-A}
	against \texttt{AC} isolates gender, since the two differ only in whether
	gender is retained.
	
	Two configurations should be explained. \texttt{PTRB} comprises
	fourteen classical voice-quality and perturbation measures---five jitter variants, six shimmer	variants, harmonics-to-noise ratio, and the mean and standard deviation of
	$F_0$---selected \emph{a priori} on clinical grounds rather than by any
	data-driven procedure, and therefore requiring no correction for selection.
	\texttt{AC-Dur} leaves acoustic features but removes phonation duration, for the reason
	given in Section~\ref{sec:audit}: we cannot determine from the released
	features whether duration reflects maximum phonation time or a fixed recording
	protocol, and no conclusion should depend on that ambiguity.
	
	\begin{table}[htb!]
		\centering
		\caption{Feature configurations. \texttt{D} denotes demographics (age and
			gender), \texttt{H} the four symptom items, \texttt{AC} the $49$ static acoustic
			descriptors, and $\Delta$ the $52$ MFCC derivative summaries. A suffix
			$-\texttt{X}$ denotes removal of \texttt{X} and $+\texttt{X}$ denotes addition of \texttt{X},  so predictor counts are additive.
			$\dagger$ marks configurations containing age.}
		\label{tab:configs}
		\small
		\setlength{\tabcolsep}{6pt}
		\begin{tabular}{@{}l| c |c@{}}
			\toprule
			Abbrev. & $p$ & Composition \\
			\midrule
			\texttt{A}\,$\dagger$              & 1   & age \\
			\texttt{D}\,$\dagger$              & 2   & age, gender \\
			\texttt{DH}\,$\dagger$             & 6   & age, gender, and symptom items \\
			\texttt{AC}                        & 49  & static acoustics \\
			\texttt{AC-Dur}                    & 48  & static acoustics (no duration) \\
			\texttt{PTRB}                      & 14  & [jitter, shimmer, HNR, $F_0$] stats \\
			\midrule
			\texttt{DAC}\,$\dagger$            & 51  & age, gender, and static acoustics \\
			\texttt{DAC+}$\Delta$\,$\dagger$   & 103 & age, gender, and (static acoustics $+\ \Delta$) \\
			\texttt{DAC-A}                     & 50  & gender and static acoustics ($-$ age)\\
			\texttt{ALL}\,$\dagger$            & 55  & all features $-\ \Delta$\\
			\texttt{ALL+}$\Delta$\,$\dagger$   & 107 & all features \\
			\texttt{ALL-A}                     & 54  & all features ($-$ age $-\ \Delta$) \\
			\bottomrule
		\end{tabular}
	\end{table}
	
	\subsection{Participant-Level Cross-Validation}
	\label{sec:cv}
	
	Let $\mathcal{P}$ denote the set of $67$ participants and $\mathcal{R}_i$ the
	recordings contributed by participant $i$, with $\lvert\mathcal{R}_i\rvert$
	ranging from $1$ to $69$ (median $8$).
	
	\subsubsection{Fold construction} Outer partitions are generated by stratified
	grouped $k$-fold method with $k=5$, grouping on participant identifier and stratifying
	on label, repeated over five seeds for $25$ outer partitions in total. All
	recordings from a participant fall on the same side of every split. Partitions
	are generated once and reused across all twelve configurations, so that any two
	configurations are evaluated on identical partitions and their scores are
	paired by construction. Hyperparameters are selected within each outer training
	set by a further grouped three-fold partition, giving a fully nested design.
	
	\subsubsection{Sample weighting} Participants contribute between one and
	$69$ recordings, thus an unweighted objective would allow a single participant to
	dominate. Each recording $r \in \mathcal{R}_i$ receives weight
	\begin{equation}
		w_r = \frac{1}{\lvert\mathcal{R}_i\rvert} \cdot
		\frac{\lvert\mathcal{R}\rvert}{\sum_{s}1/\lvert\mathcal{R}_{i(s)}\rvert},
		\label{eq:weights}
	\end{equation}
	so that every participant contributes equal total weight and the mean weight is
	one. Weights are applied to the training objective only. Evaluation is
	unweighted, since aggregation already reduces each participant to a single
	score.
	
	\subsubsection{Score aggregation} A model returns one score $p_r$ per recording,
	but the quantity of interest is one score $\hat{p}_i$ per participant. We aggregate in the logit domain,
	\begin{equation}
		\hat{p}_i = \sigma\!\left(
		\frac{1}{\lvert\mathcal{R}_i\rvert}\sum_{r \in \mathcal{R}_i}
		\log \frac{p_r}{1-p_r} \right),
		\label{eq:agg}
	\end{equation}
	where $\sigma(\cdot)$ denotes the sigmoid function, and $p_r$ clipped to $[\epsilon, 1-\epsilon]$, $\epsilon = 10^{-12}$. Logit averaging corresponds to a product-of-experts pooling of the per-recording evidence and is the natural choice when recordings are treated as repeated	measurements of one underlying state. It is, however, sensitive to individual confident predictions, and we verified that substituting the arithmetic mean of scores does not alter the ordering of configurations.
	
	\subsubsection{Pooled scores} Each participant is held out in exactly one fold per
	repetition, yielding five out-of-sample scores. These are averaged to give one
	pooled score per participant per configuration, from which all reported metrics
	are computed. 
	
	\subsection{Model and Hyperparameter Selection}
	\label{sec:model}
	
	We use gradient-boosted decision trees (CatBoost~\cite{prokhorenkova2018catboost}),
	matching the model family reported to perform best on this corpus in prior
	work~\cite{idrisoglu2024copdvd}, with gender handled natively as a categorical
	feature. The grid comprises $4\times3\times4\times4 = 192$ combinations of the hyperparameters presented in Table~\ref{tab:hp}.
	\begin{table}[htbp]
		\centering
		\caption{CatBoost hyperparameter search space.}
		\label{tab:hp}
		\small
		\begin{tabular}{ll}
			\hline
			\textbf{Hyperparameter} & \textbf{Values} \\
			\hline
			Iterations                 & $\{300, 600, 800, 1000\}$ \\
			Tree depth                 & $\{2, 4, 6\}$ \\
			Learning rate              & $\{0.001, 0.01, 0.03, 0.10\}$ \\
			$L_2$ leaf regularization  & $\{1, 3, 6, 10\}$ \\
			\hline
		\end{tabular}
	\end{table}
	
	Selection uses a one-standard-error rule rather than the maximizer of inner
	cross-validated ROC-AUC~\cite{hastie2009elements}. Writing $\bar{a}_g$ and
	$\mathrm{SE}(a_g)$ for the mean and standard error of subject-level AUC across
	inner folds, respectively, for grid point $g$, and $g^\star = \arg\max_g \bar{a}_g$, we select
	\begin{equation}
		\hat{g} = \arg\min_{\,g \,:\, \bar{a}_g \,\geq\, \bar{a}_{g^\star} - \mathrm{SE}(a_{g^\star})}
		\ \mathrm{complexity}(g),
		\label{eq:onese}
	\end{equation}
	where complexity orders grid points by depth, then iterations, then by
	\emph{decreasing} $L_2$ penalty (so that stronger regularization is
	preferred), then learning rate.
	
	Using such a rule can be justified. Inner folds contain approximately $18$ participants, so inner AUC is estimated with a standard error of roughly $0.10$, while adjacent grid points differ by far less. The maximizer is therefore highly susceptible to selection on noise. In preliminary runs we	observed that this instability moved pooled ROC-AUC by up to $0.08$ for
	one- and two-predictor configurations between otherwise identical executions,
	while configurations with $50$ or more predictors reproduced within $0.02$.
	The pattern is expected: with many predictors the data constrain the fit and
	the grid point matters little, whereas with a single predictor the choice of
	depth and iteration count determines the shape of the fitted step function in
	age. The demographic baselines are the reference against which acoustic
	performance is judged, hence an unstable baseline is not an acceptable starting
	point. 
	
	\subsection{Confidence Intervals}
	\label{sec:ci}
	
	All resampling is over participants. Resampling recordings would treat
	repeated measurements of one person as independent observations and yield
	intervals that are too narrow. Marginal intervals for ROC-AUC and average precision use $2000$ bootstrap resamples of the $67$ pooled participant scores, reported as $2.5$th and
	$97.5$th percentiles. Because average precision has a chance level equal to
	class prevalence, we report that baseline alongside every PR-AUC and flag
	whether the interval excludes it. Comparisons between configurations use a paired, class-stratified bootstrap with $20{,}000$ resamples. Cases and controls are resampled separately to preserve class balance, and both configurations are evaluated on the same
	resampled participants, so that the sampling variation common to both cancels.
	Reporting the difference of two marginal intervals would substantially
	overstate uncertainty here, since the configurations are nested and their
	scores are strongly correlated.
	
	\subsection{Matched-Cohort Evaluation}
	\label{sec:matched}
	
	Section~\ref{sec:audit} established that a non-monotone function of age alone
	attains ROC-AUC $0.721$ on this cohort. Pooled evaluation therefore cannot
	distinguish a model that has learned acoustic pathology from one that has
	learned age. The matched-cohort procedure evaluates the same fitted models on
	subsets in which age discrimination is strongly reduced and directly quantified.
	
	\paragraph{Cohort construction} For each of $2000$ repetitions we draw a
	one-to-one age-matched cohort. COPD participants are processed in random order, which means that each is matched to the nearest not-yet-used control within a caliper of \textit{two} years, and unmatched participants are discarded. This yields a mean of
	approximately $24$ pairs from the $67$ participants, the excluded participants
	being predominantly the controls with no age counterpart identified in
	Fig.~\ref{fig:age}(b). Matching is one-to-one, thus the matched cohorts have
	prevalence exactly $0.5$, and the chance level for average precision is
	therefore $0.5$ rather than the cohort value of $0.448$.
	
	Two properties of this construction matter. First, all twelve configurations are
	scored on the \emph{same} matched cohort within each repetition, so
	between-configuration differences are paired and not attributable to different
	draws. Second, matching is applied at evaluation time to models trained on the
	unmatched cohort. This tests whether the signal a model has learned is still predictive enough after the removal of age information. 
	
	\paragraph{Verified controls} On every matched cohort we compute the
	discrimination achieved by raw (unmodeled) age and by raw gender, and report both. This is the component we consider most important methodologically. Matching is
	conventionally verified through a balance statistic such as the SMD of Eq.~\eqref{eq:smd}, but the latter is a difference of means whereas
	AUC is a rank statistic, and the two can disagree: distributions with equal
	means and unequal shape yield SMD near zero while remaining rank-separable. We therefore report the AUC of the confounder itself on the cohort used for
	evaluation. Only cohorts on which raw age discriminates at chance support an
	age-controlled reading of the remaining results, and this condition is checked
	before any other matched quantity is interpreted.
	
	\paragraph{Interval estimates} Percentiles taken across matching repetitions
	describe only the variability induced by which cohort was drawn, with the
	participants held fixed, and are not confidence intervals. For the primary
	analysis we therefore draw, within each repetition, one bootstrap resample of
	the matched cohort before scoring, so that the resulting percentiles reflect
	both matching variability and participant sampling. That resample is over \emph{matched pairs}, not over cases and controls
	separately. The pair is the sampling unit the design creates, and resampling
	the two classes independently preserves age balance only in expectation: a case
	may be drawn without its control, so the class marginals agree on average while
	individual replicates need not. Under class-independent resampling on this
	cohort, $73\,\%$ of replicates exceeded $\lvert\mathrm{SMD}\rvert = 0.1$,
	$48\,\%$ exceeded $0.2$, and in $14\,\%$ raw age discriminated above $0.60$, so
	a substantial fraction of the cohorts being scored were not in fact age-matched
	and the verification above would have held only on average. Resampling pairs
	preserves balance in every replicate ($2.4\,\%$ above $0.1$, none above $0.2$).
	Its effect on the intervals is itself diagnostic and is reported in
	Section~\ref{sec:res-matched}. 
	
	\subsection{Feature Importance and Uncertainty Quantification}
	\label{sec:aux}
	
	Permutation importance is computed at participant level. Permuting rows
	independently would destroy the within-participant correlation structure and
	inflate apparent importance; instead we pair each participant with a randomly
	chosen donor participant and resample the feature of interest from the donor's
	recordings, preserving the repeated-measures structure. Importance is the
	resulting drop in subject-level AUC, averaged over five repetitions per outer
	split. Because rankings from small samples are unstable, we also report the
	mean Spearman correlation between importance rankings across outer splits as a
	measure of how far any feature-level claim can be taken.
	
	As an exploratory assessment of deployment behaviour we compute conformal prediction sets~\cite{vovk2005algorithmic} at $\alpha \in \{0.10, 0.20\}$. The	construction is stated in full, because the validity that may be claimed depends on it. The exchangeable unit is the participant: recording-level scores are	aggregated by Eq.~\eqref{eq:agg} to one score per participant before any	calibration quantity is formed. Within each outer training set we generate a label-stratified, participant-grouped partition into $K = \min(5, \lvert\mathcal{P}_{\mathrm{train}}\rvert)$ folds. Each fold is scored by a model fitted on the remaining $K-1$, giving one out-of-fold score per training
	participant, and the nonconformity score is 
	\begin{equation}
		s_i = \lvert y_i - \hat{p}_i\rvert
	\end{equation} 
	The $n$ scores are \emph{pooled into a single calibration set} and one threshold $q_\alpha$ is taken as the $\lceil (n+1)(1-\alpha)\rceil$-th order statistic, that is, fold-wise thresholds are neither formed nor aggregated. A single model is then refitted on the entire outer training set and applied to the test participants, and the prediction set for a test participant admits label $0$ when $\hat{p} \leq q_\alpha$ and label $1$ when $\hat{p} \geq 1 - q_\alpha$. Because one threshold is taken over both
	classes jointly, the target is marginal rather than class-conditional coverage.
	We report coverage as an empirical quantity and treat the analysis as
	descriptive of deployment behaviour rather than as an inferential claim. Sets
	may be empty, singleton, or contain both labels; we report coverage together
	with the abstention rate, since a set containing both labels covers the truth
	trivially and coverage alone is consequently uninformative.
	
	\subsection{Independent Replication}
	\label{sec:replication}
	
	To establish that the findings are properties of the data rather than of one
	model family or of the selection procedure, we repeat the entire protocol with
	two further learners: histogram-based gradient boosting~\cite{pedregosa2011scikit, ke2017lightgbm} and $L_2$-regularized logistic regression on standardized features~\cite{hastie2009elements}. Both use fixed hyperparameters with no inner cross-validation, which removes selection instability entirely.
	Fold construction, weighting, aggregation, and matched evaluation---including
	the paired bootstrap over matched pairs---are unchanged, and we report the
	matched result for both. The logistic model additionally serves as a linearity
	probe: a substantial gap between the learners indicates that the
	acoustic--disease relationship is not well approximated by a linear decision
	boundary.

	\section{Results}
	\label{sec:results}
	
	We present the conventional pooled analysis first (Section~\ref{sec:res-pooled}), because its inability to answer the question of	interest is itself a result. The matched-cohort analysis follows (Section~\ref{sec:res-matched}), and	Section~\ref{sec:res-repl} establishes that the findings do not depend on the
	choice of learner.
	
	\subsection{Pooled Evaluation}
	\label{sec:res-pooled}
	
		\begin{table*}[htb!]
		\centering
		\caption{Pooled subject-level discrimination, all $67$ participants, with
			participant bootstrap intervals ($2000$ resamples). Chance is $0.5$ for ROC-AUC
			and the prevalence $0.448$ for PR-AUC. $\dagger$ marks configurations containing
			the explicit age predictor.}
		\label{tab:pooled}
		\small
		\setlength{\tabcolsep}{8pt}
		\begin{tabular}{@{}l r l l@{}}
			\toprule
			Configuration & $p$ & ROC-AUC & PR-AUC \\
			\midrule
			\texttt{DAC+}$\Delta$\,$\dagger$   & 103 & 0.741 [0.622, 0.855] & 0.697 [0.540, 0.864] \\
			\texttt{ALL+}$\Delta$\,$\dagger$        & 107 & 0.717 [0.593, 0.840] & 0.669 [0.509, 0.858] \\
			\texttt{D}\,$\dagger$                 & 2   & 0.716 [0.580, 0.844] & 0.657 [0.498, 0.859] \\
			\texttt{DAC}\,$\dagger$            & 51  & 0.712 [0.576, 0.835] & 0.692 [0.535, 0.865] \\
			\texttt{A}\,$\dagger$             & 1   & 0.709 [0.571, 0.840] & 0.596 [0.436, 0.803]\rlap{$^{*}$} \\
			\texttt{DH}\,$\dagger$        & 6   & 0.704 [0.565, 0.833] & 0.618 [0.462, 0.830] \\
			\texttt{ALL-A}                    & 54  & 0.703 [0.572, 0.827] & 0.692 [0.524, 0.849] \\
			\texttt{DAC-A}               & 50  & 0.702 [0.570, 0.830] & 0.674 [0.508, 0.849] \\
			\texttt{AC}                       & 49  & 0.696 [0.567, 0.820] & 0.677 [0.507, 0.849] \\
			\texttt{AC-Dur}              & 48  & 0.691 [0.556, 0.818] & 0.698 [0.544, 0.837] \\
			\texttt{ALL}\,$\dagger$                 & 55  & 0.677 [0.541, 0.807] & 0.657 [0.487, 0.836] \\
			\texttt{PTRB}                    & 14  & 0.671 [0.537, 0.799] & 0.703 [0.541, 0.835] \\
			\bottomrule
		\end{tabular}
		\\[2pt]
		\raggedright\footnotesize $^{*}$ the only interval that does not exclude the
		PR-AUC chance baseline.
	\end{table*}
	
	Table~\ref{tab:pooled} reports pooled subject-level discrimination for all
	twelve configurations. \emph{Every} configuration has a ROC-AUC interval
	excluding chance, and eleven of twelve exceed the prevalence baseline on
	average precision. This includes \texttt{A}, a single-predictor model,
	at ROC-AUC $0.709$ $[0.571, 0.840]$.
	
	The estimates span $0.671$--$0.741$, a range of $0.070$, against interval widths
	of $0.233$--$0.268$. A one-predictor age model
	($0.709$) is therefore statistically indistinguishable from a $103$-predictor
	combined model ($0.741$), and both are indistinguishable from every other
	configuration. The ordering is not interpretable: \texttt{D}, with age
	and gender alone, ranks third of twelve and above every acoustics-only variant.
	
	This is the expected consequence of the cohort structure documented in 	Section~\ref{sec:audit}. A hand-specified non-monotone function of age attains $0.721$ on this cohort (Fig.~\ref{fig:nonmono}), which bounds from below what any learner with access to age can achieve without using acoustic information at all. Pooled evaluation cannot distinguish a model that has learned respiratory pathology from one that has learned the recruitment age	distribution.
	
	One asymmetry between the two metrics is worth noting. On average precision the
	ordering partially inverts: \texttt{PTRB} ranks first ($0.703$)	and \texttt{A} last ($0.596$), and \texttt{A} is the sole configuration whose interval fails to exclude the prevalence baseline. Average precision weights the positive class more heavily than ROC-AUC and is correspondingly more sensitive to the difference between the two model families, even before matching.
	
	Table~\ref{tab:paired} reports the sixteen paired comparisons. \emph{None}
	excludes zero, on either metric. The comparisons differ substantially in precision, and this distinction matters more than the point estimates. Two are narrow enough to constitute resolved	null results: adding the symptom-questionnaire block to the age-free acoustic model changes ROC-AUC by $+0.001$ $[-0.040, +0.037]$, and removing phonation duration changes it by $+0.005$ $[-0.040, +0.049]$. Gender is moderately tight
	at $+0.005$ $[-0.068, +0.087]$. Three predictor blocks in routine use therefore
	contribute nothing detectable.
	
	The remaining comparisons are underpowered rather than null. The incremental
	value of age is $-0.026$ $[-0.151, +0.095]$ and $+0.010$ $[-0.098, +0.117]$,
	acoustics versus demographics alone ranges from $-0.045$ to $+0.025$ with
	interval widths of $0.26$--$0.42$. No conclusion about either can be drawn from
	the pooled design.

	\begin{table*}[htb!]
		\centering
		\caption{Paired comparisons on pooled scores, class-stratified participant
			bootstrap ($20{,}000$ resamples). No interval excludes zero. Ordered by interval
			width; only the first two are narrow enough to constitute resolved nulls.}
		\label{tab:paired}
		\small
		\setlength{\tabcolsep}{8pt}
		\begin{tabular}{@{}l l l r@{}}
			\toprule
			Contrast & Isolates & $\Delta$ROC-AUC & width \\
			\midrule
			\texttt{ALL-A} $-$ \texttt{DAC-A} & health items & $+0.001$ $[-0.040, +0.037]$ & 0.077 \\
			\texttt{AC} $-$ \texttt{AC-Dur} & duration & $+0.005$ $[-0.040, +0.049]$ & 0.088 \\
			\texttt{ALL} $-$ \texttt{ALL+}$\Delta$ & MFCC $\Delta$ & $-0.041$ $[-0.103, +0.014]$ & 0.116 \\
			\texttt{DH} $-$ \texttt{D} & health items & $-0.013$ $[-0.077, +0.044]$ & 0.121 \\
			\texttt{DAC} $-$ \texttt{DAC+}$\Delta$ & MFCC $\Delta$ & $-0.030$ $[-0.104, +0.038]$ & 0.141 \\
			\texttt{DAC-A} $-$ \texttt{AC} & gender & $+0.005$ $[-0.068, +0.087]$ & 0.156 \\
			\texttt{PTRB} $-$ \texttt{AC} & parsimony & $-0.025$ $[-0.113, +0.064]$ & 0.177 \\
			\texttt{DAC} $-$ \texttt{DAC-A} & age & $+0.010$ $[-0.098, +0.117]$ & 0.215 \\
			\texttt{ALL} $-$ \texttt{ALL-A} & age & $-0.026$ $[-0.151, +0.095]$ & 0.247 \\
			\texttt{DAC+}$\Delta$ $-$ \texttt{D} & acoustics & $+0.025$ $[-0.107, +0.157]$ & 0.264 \\
			\texttt{ALL+}$\Delta$ $-$ \texttt{D} & acoustics & $+0.001$ $[-0.134, +0.135]$ & 0.269 \\
			\texttt{ALL} $-$ \texttt{D} & acoustics & $-0.040$ $[-0.186, +0.103]$ & 0.289 \\
			\texttt{PTRB} $-$ \texttt{ALL} & parsimony & $-0.005$ $[-0.149, +0.143]$ & 0.292 \\
			\texttt{DAC} $-$ \texttt{D} & acoustics & $-0.005$ $[-0.153, +0.140]$ & 0.293 \\
			\texttt{PTRB} $-$ \texttt{D} & acoustics & $-0.045$ $[-0.232, +0.141]$ & 0.373 \\
			\texttt{AC} $-$ \texttt{D} & acoustics & $-0.020$ $[-0.209, +0.167]$ & 0.376 \\
			\bottomrule
		\end{tabular}
	\end{table*}
	
	\subsection{Matched-Cohort Evaluation}
	\label{sec:res-matched}
	
	Under matched-cohort evaluation and at the primary caliper of two years (Table~\ref{tab:matched}), the achieved participant-level age SMD is $0.026$, down from $0.726$	in the unmatched cohort, with fewer than $3\,\%$ of replicates exceeding
	$\lvert\mathrm{SMD}\rvert = 0.1$. Raw age on the matched cohorts discriminates
	at $0.510$ $[0.469, 0.551]$ and raw gender at $0.479$, that is, both are at chance level. 
	
		\begin{table*}[htb!]
		\centering
		\caption{Matched-cohort discrimination, caliper $2$ years, $2000$ repetitions
			with a paired bootstrap over matched pairs. Chance is $0.5$ for both metrics
			under one-to-one matching. Raw age scores $0.510$ $[0.469, 0.551]$ and raw gender $0.479$ on these cohorts. Bold lower bounds exclude chance. $\delta$ is the matched-subset AUC minus full-cohort pooled AUC. $\dagger$ contains the explicit age predictor.}
		\label{tab:matched}
		\small
		\setlength{\tabcolsep}{8pt}
		\begin{tabular}{@{}l r l l c@{}}
			\toprule
			Configuration & $p$ & Matched ROC-AUC & Matched PR-AUC & $\delta$ \\
			\midrule
			\texttt{AC}                      &  49 & 0.717 [\textbf{0.552}, 0.859] & 0.747 [\textbf{0.581}, 0.892] & $+0.021$ \\
			\texttt{DAC-A}                   &  50 & 0.682 [0.498, 0.845]          & 0.734 [\textbf{0.566}, 0.888] & $-0.020$ \\
			\texttt{AC-Dur}                  &  48 & 0.681 [\textbf{0.505}, 0.835] & 0.740 [\textbf{0.586}, 0.869] & $-0.010$ \\
			\texttt{DAC+}$\Delta$\,$\dagger$ & 103 & 0.679 [\textbf{0.524}, 0.818] & 0.703 [\textbf{0.541}, 0.851] & $-0.062$ \\
			\texttt{ALL-A}                   &  54 & 0.679 [0.495, 0.842]          & 0.741 [\textbf{0.567}, 0.887] & $-0.024$ \\
			\texttt{PTRB}                    &  14 & 0.665 [0.488, 0.823]          & \textbf{0.755} [\textbf{0.612}, 0.873] & $-0.006$ \\
			\texttt{ALL+}$\Delta$\,$\dagger$ & 107 & 0.632 [0.479, 0.769]          & 0.668 [\textbf{0.512}, 0.822] & $-0.085$ \\
			\texttt{DAC}\,$\dagger$          &  51 & 0.626 [0.474, 0.778]          & 0.695 [\textbf{0.538}, 0.848] & $-0.086$ \\
			\texttt{D}\,$\dagger$            &   2 & 0.595 [0.410, 0.760]          & 0.629 [0.496, 0.811]          & $-0.121$ \\
			\texttt{ALL}\,$\dagger$          &  55 & 0.580 [0.431, 0.736]          & 0.640 [0.493, 0.799]          & $-0.097$ \\
			\texttt{DH}\,$\dagger$           &   6 & 0.575 [0.394, 0.747]          & 0.586 [0.461, 0.764]          & $-0.129$ \\
			\texttt{A}\,$\dagger$            &   1 & 0.531 [0.423, 0.649]          & 0.534 [0.463, 0.647]          & $-0.178$ \\
			\bottomrule
		\end{tabular}
	\end{table*}
	
	The effect of resampling pairs rather than classes (Section~\ref{sec:matched})
	is itself diagnostic, because the two model families respond in opposite
	directions. Configurations containing age \emph{narrow}---the width of the
	interval for \texttt{A}, a near-pure function of age, falls from $0.337$ to
	$0.227$---having previously inherited variance from replicates in which the
	matching had been destroyed. Configurations excluding age \emph{widen} slightly,
	from a mean width of $0.318$ to $0.333$, correctly reflecting that $24$ pairs
	and not $48$ independent participants are the units of resampling. The
	per-replicate variance of the raw-age control falls by a factor of $14.8$.
	
	Table~\ref{tab:matched} reports discrimination on these cohorts. The two model
	families separate on point estimates: configurations excluding age span
	$0.665$--$0.717$ ROC-AUC against $0.531$--$0.679$ for those containing age, with
	only \texttt{DAC+}$\Delta$ overlapping. The separation is unambiguous on average
	precision, where all five age-free configurations exclude the matched chance
	level of $0.5$ and only three of seven age-containing configurations do. On
	ROC-AUC, the marginal intervals are less decisive: two of five age-free
	configurations exclude chance, against one of seven age-containing. At $24$
	pairs the standard error of an AUC is approximately $0.076$, so a marginal
	against-chance interval is an insensitive instrument here (\texttt{DAC-A} and
	\texttt{ALL-A} miss the bar by $0.002$ and $0.005$) and we do not rest the
	claim on it.
	
	The change induced by matching is the more stable summary. Averaged within
	groups, matching shifts age-free configurations by $-0.008$ ROC-AUC and
	age-containing configurations by $-0.108$. The age-free models are stable under restriction to repeatedly drawn age-matched subsets.
	
	Two features of Table~\ref{tab:matched} qualify a purely binary reading. First,
	\texttt{DAC+}$\Delta$ contains age and nonetheless clears the ROC chance bar, at
	a lower bound of $0.524$. Second, and relatedly, the magnitude of the
	matching-induced drop among age-containing configurations scales inversely with
	the number of acoustic predictors available. The twelve configurations form two
	nested ladders, and the relation is monotone along each:
	\texttt{A}~($p{=}1$, $-0.178$)~$\rightarrow$~\texttt{D}~($2$, $-0.121$)
	$\rightarrow$~\texttt{DAC}~($51$, $-0.086$)
	$\rightarrow$~\texttt{DAC+}$\Delta$~($103$, $-0.062$), and
	\texttt{A}~($1$, $-0.178$)~$\rightarrow$~\texttt{DH}~($6$, $-0.129$)
	$\rightarrow$~\texttt{ALL}~($55$, $-0.097$)
	$\rightarrow$~\texttt{ALL+}$\Delta$~($107$, $-0.085$). The ordering holds within
	each ladder but not between them, so the governing quantity is the acoustic
	information available relative to the same demographic block, not the raw
	predictor count. The effect is graded rather than dichotomous, and the gradient
	is informative: when a confounded covariate is available and acoustic
	information is scarce, the model relies on the covariate: as acoustic
	information increases, it has an alternative and the reliance diminishes.
	Supplying age at training time does not leave the acoustic representation
	intact.
	
	An important outcome of the analysis is on average precision. Under matching cohorts, matched prevalence is exactly $0.5$ (chance level) in every replicate under paired resampling. The \texttt{PTRB} set, fourteen \emph{a priori} clinical measures, achieves PR-AUC $0.755$ $[0.612, 0.873]$ (the highest lower bound of any configuration under this learner) and \texttt{AC} achieves $0.747$ $[0.581, 0.892]$.
	
	However, an important note should be stressed. Of the twelve paired contrasts against raw age	on the same matched cohorts, only \texttt{AC} excludes zero, at $+0.207$
	$[+0.027, +0.365]$. A single exclusion across twelve comparisons without
	multiplicity correction is suggestive and no more, although the five age-free
	configurations occupy five of the six largest contrasts.
	
	\subsection{Independent Replication}
	\label{sec:res-repl}
	
	Repeating the protocol with histogram-based gradient boosting under fixed
	hyperparameters---no inner cross-validation, and therefore no exposure to
	selection instability---reproduces the matched-cohort dissociation.
	Table~\ref{tab:repl} gives the comparison. The two learners agree closely at the
	level of the family contrast and less so within families: the mean matched
	ROC-AUC of the age-free configurations is $0.685$ under CatBoost and $0.702$
	under histogram boosting, while the age-containing means are $0.603$ and
	$0.556$. Across the twelve configurations, CatBoost and HistGB showed strong agreement, with Pearson $r=0.902$ and Spearman $\rho=0.935$, and a mean absolute difference of $0.035$. 
	
	By the family contrast the replication is in fact cleaner. Under histogram-based
	boosting the age-containing configurations average $0.556$ and
	none of the seven clears the ROC bar, while all five age-free
	configurations do, against two of five and one of seven under CatBoost.
	\texttt{A} falls to $0.502$ $[0.351, 0.665]$. The highest PR-AUC lower bound in
	the entire study is \texttt{PTRB} under this learner, at $0.779$
	$[0.651, 0.888]$. 
	
	$L_2$-regularised logistic regression behaves informatively in three ways.
	First, it shows that the pooled comparison is confounded even under a linear
	decision boundary. Pooled (not shown in Table~\ref{tab:repl}), age outperforms acoustics ($0.668$ versus $0.595$) while on matched cohorts the ordering reverses ($0.497$ versus $0.583$), as it does under both boosted ensembles. The direction of the dissociation is therefore common to all three learners: age-free minus age-containing family means of $+0.082$, $+0.147$, and $+0.066$ for CatBoost, HistGB, and logistic regression, respectively, although the linear model is underpowered: no configuration clears the chance bar and its intervals are approximately $0.37$ wide.
	
	Second, the linear model gives the sharpest evidence that a fitted age-only
	model is not a valid control. Under logistic regression, exactly three
	configurations have paired contrasts against raw age whose intervals exclude
	zero, all of them negative, and they are precisely the three age-dominant ones:
	\texttt{A} at $-0.043$ $[-0.091, -0.005]$, \texttt{D} at $-0.077$
	$[-0.150, -0.015]$, and \texttt{DH} at $-0.110$ $[-0.222, -0.010]$. No other
	configuration under any learner shows this. 
	
	Third, the gap between the linear and boosted learners on the pooled acoustic
	configurations is consistent with the acoustic?disease relationship not being well approximated linearly. We note that the comparison confounds model family with
	regularization strength, since the penalty was fixed rather than tuned.
	
	\begin{table}[t]
		\centering
		\caption{Matched ROC-AUC ($2$-year caliper, paired bootstrap over matched
			pairs) under CatBoost with tuned hyperparameters, histogram-based gradient
			boosting with fixed hyperparameters, and $L_2$-regularised logistic
			regression. Bold indicates the lower bound excludes chance.}
		\label{tab:repl}
		\small
		\setlength{\tabcolsep}{7pt}
		\begin{tabular}{@{}l c c c@{}}
			\toprule
			Configuration & CatBoost & HistGB & LogReg \\
			\midrule
			\texttt{AC}                        & \textbf{0.717} & \textbf{0.717} & 0.583 \\
			\texttt{DAC-A}                     & 0.682 & \textbf{0.710} & 0.549 \\
			\texttt{AC-Dur}                    & \textbf{0.681} & \textbf{0.724} & 0.606 \\
			\texttt{DAC+}$\Delta$\,$\dagger$   & \textbf{0.679} & 0.595 & 0.597 \\
			\texttt{ALL-A}                     & 0.679 & \textbf{0.693} & 0.545 \\
			\texttt{PTRB}                      & 0.665 & \textbf{0.668} & 0.628 \\
			\texttt{ALL+}$\Delta$\,$\dagger$   & 0.632 & 0.596 & 0.573 \\
			\texttt{DAC}\,$\dagger$            & 0.626 & 0.576 & 0.580 \\
			\texttt{D}\,$\dagger$              & 0.595 & 0.554 & 0.434 \\
			\texttt{ALL}\,$\dagger$            & 0.580 & 0.569 & 0.549 \\
			\texttt{DH}\,$\dagger$             & 0.575 & 0.497 & 0.382 \\
			\texttt{A}\,$\dagger$              & 0.531 & 0.502 & 0.497 \\
			\midrule
			\footnotesize Age-free clearing: & $2/5$ & $5/5$ & $0/5$\\
			\footnotesize Age-containing clearing: & $1/7$ & $0/7$ & $0/7$\\
			\footnotesize Family mean, age-free: & $0.685$ & $0.702$ & $0.582$\\
			\footnotesize Family mean, age-containing: &$0.603$ &$0.556$ & $0.516$\\
			\bottomrule
		\end{tabular}
\end{table}
	
	\subsection{Feature Importance and Conformal Prediction}
	\label{sec:res-aux}
	
	Permutation importance was computed for three configurations at participant
	level, permuting each feature between randomly paired participants so that the
	within-participant correlation structure is preserved.
	
	The results are dominated by a single descriptor. The standard deviation of
	$F_0$ ranks first in both age-free configurations analyzed (\texttt{DAC-A}, mean AUC drop $0.040$; \texttt{PTRB}, $0.071$) and second in \texttt{ALL}
	($0.028$). Importance is extremely concentrated: in the $50$-predictor age-free
	configuration, only five features have positive mean importance at all and $F_0$
	variability accounts for $94\,\%$ of the total; in the $14$-predictor clinical
	set the figure is $60\,\%$. That $F_0$ variability should carry the signal is
	physiologically coherent---unstable fundamental frequency reflects irregular
	vocal-fold vibration under reduced and fluctuating subglottal pressure---but
	we emphasise below that this analysis cannot establish it.
	
	In \texttt{ALL}, which contains the explicit age predictor, age ranks
	first with an importance of $0.060$, more than double that of the highest-ranked
	acoustic feature. This is a direct measurement of the mechanism inferred in
	Section~\ref{sec:res-matched}: when a confounded covariate is supplied, the
	model preferentially exploits it, and the acoustic descriptors are relegated to
	a secondary role.
	
	The corresponding rank stability is poor and we report it as a limit on
	interpretation rather than a supporting result. Mean Spearman correlation
	between importance rankings across the $25$ outer splits is $0.032$ for the
	$50$-predictor configuration, $0.111$ for the $55$-predictor configuration, and
	$0.202$ for the $14$-predictor clinical set (standard deviations
	$\approx 0.25$ throughout, $300$ split pairs each). Rankings therefore do not
	replicate across resamplings of a cohort this size, and no claim about the
	relative importance of individual features is warranted. Stability improves
	monotonically as dimensionality falls, which is the expected direction and a
	further argument for compact feature sets. The one observation that does not
	depend on within-configuration rank stability is that $F_0$ variability emerges
	as the leading acoustic descriptor in three independently constructed feature
	sets.
		\begin{table*}[t]
		\centering
		\caption{Exploratory cross-conformal prediction sets, averaged over $25$ outer
			splits. Observed coverage met or exceeded the nominal level $1-\alpha$
			throughout; the construction carries no finite-sample guarantee
			(Section~\ref{sec:aux}). No prediction set was empty, so abstention consists
			entirely of sets containing both labels. ``Efficiency'' is the proportion of
			participants receiving a single-label set, ``Correctness'' the accuracy among
			those, and $\mathrm{E}\!\times\!\mathrm{C}$ their product---the proportion of
			all participants receiving a single label and correct prediction.
			$\dagger$ contains the explicit age predictor.}
		\label{tab:conformal}
		\small
		\setlength{\tabcolsep}{5pt}
		\begin{tabular}{@{}l cccc cccc@{}}
			\toprule
			& \multicolumn{4}{c}{$\alpha=0.10$} & \multicolumn{4}{c}{$\alpha=0.20$} \\
			\cmidrule(lr){2-5}\cmidrule(lr){6-9}
			Configuration & Coverage & Efficiency & Correctness & E$\times$C
			& Coverage & Efficiency & Correctness & E$\times$C \\
			\midrule
			\texttt{PTRB}                   & 0.913 & 0.387 & 0.798 & 0.309 & 0.808 & 0.737 & 0.752 & 0.554 \\
			\texttt{AC}                     & 0.920 & 0.261 & 0.705 & 0.184 & 0.814 & 0.675 & 0.737 & 0.498 \\
			\texttt{AC-Dur}                 & 0.925 & 0.252 & 0.706 & 0.178 & 0.805 & 0.697 & 0.727 & 0.507 \\
			\texttt{DAC-A}                  & 0.928 & 0.269 & 0.748 & 0.201 & 0.821 & 0.693 & 0.756 & 0.524 \\
			\texttt{ALL-A}                  & 0.925 & 0.284 & 0.754 & 0.214 & 0.823 & 0.656 & 0.742 & 0.487 \\
			\texttt{ALL}\,$\dagger$         & 0.943 & 0.291 & 0.843 & 0.245 & 0.830 & 0.650 & 0.753 & 0.490 \\
			\texttt{DAC}+$\Delta$\,$\dagger$ & 0.939 & 0.263 & 0.809 & 0.213 & 0.836 & 0.621 & 0.749 & 0.466 \\
			\texttt{DAC}\,$\dagger$         & 0.951 & 0.256 & 0.861 & 0.221 & 0.839 & 0.629 & 0.767 & 0.482 \\
			\texttt{ALL}+$\Delta$\,$\dagger$ & 0.936 & 0.261 & 0.813 & 0.213 & 0.847 & 0.589 & 0.762 & 0.449 \\
			\texttt{DH}\,$\dagger$          & 0.949 & 0.219 & 0.808 & 0.177 & 0.845 & 0.461 & 0.707 & 0.326 \\
			\texttt{A}\,$\dagger$           & 0.955 & 0.235 & 0.845 & 0.199 & 0.875 & 0.449 & 0.758 & 0.340 \\
			\texttt{D}\,$\dagger$           & 0.946 & 0.207 & 0.792 & 0.164 & 0.858 & 0.429 & 0.711 & 0.305 \\
			\bottomrule
		\end{tabular}
	\end{table*}

	Table~\ref{tab:conformal} reports the exploratory cross-conformal analysis.	Observed coverage met or exceeded the nominal level for every configuration at both risk levels in this experiment. As set out in Section~\ref{sec:aux}, the construction carries no finite-sample guarantee, so this is an empirical statement and not a validity claim. It is worth mentioning that prediction sets are never empty: all abstention consists of sets containing both labels.
	
	The efficiency (single-label set output) figures are the substantive result. At $\alpha = 0.10$, nominal $90\,\%$ coverage is achieved only by declining to commit on $61$--$79\,\%$ of participants. Relaxing to $\alpha = 0.20$ reduces	abstention to $26$--$57\,\%$. Reporting coverage without abstention would	therefore substantially misrepresent what such a system delivers, and no configuration examined here approaches the efficiency that a screening instrument would require.
	
	Two patterns within Table~\ref{tab:conformal} are informative. First, at $\alpha = 0.20$ abstention separates the two model families without	overlap: $0.263$--$0.344$ for the five age-free configurations against $0.350$--$0.571$ for the seven containing age. Thus, the dissociation of Section~\ref{sec:res-matched} recurs in a quantity computed without any matching procedure. Second, correctness (accuracy among decisive predictions) must be read jointly with how often a decision is issued, since a model that commits only on unambiguous cases will appear more accurate while deciding less often. At $\alpha = 0.10$ the age-containing configurations are the more accurate when they commit	($0.792$--$0.861$ against $0.705$--$0.798$) but commit least often
	($0.207$--$0.291$ against $0.252$--$0.387$), which means the two effects largely cancel. The	product of the two, the proportion of all participants receiving a decisive and
	correct prediction, is the quantity that does not admit this trade-off. At
	$\alpha = 0.20$ it reaches $0.487$--$0.554$ for the five age-free configurations
	and falls to $0.305$--$0.340$ for the three demographic-only configurations
	\texttt{A}, \texttt{D} and \texttt{DH}. We note that it does not separate the
	two families cleanly (\texttt{ALL} attains $0.490$, marginally above
	\texttt{ALL-A} at $0.487$) so the decisive fraction, which does separate them
	without overlap, is the sharper statistic here. \texttt{PTRB} is the most
	efficient configuration on both measures and at both risk levels, committing on
	$73.7\,\%$ of participants at $\alpha = 0.20$ and returning a decisive, correct
	prediction for $55.4\,\%$ of them.

	\section{Discussion}
	\label{sec:discussion}
	The matched-cohort analysis and its replication under two additional learners converge on the same qualitative conclusion. On matched cohorts where raw age and raw gender are both measured to discriminate at chance, configurations excluding age retain ROC-AUC $0.665$--$0.717$ and average precision $0.734$--$0.755$ against a one-to-one baseline of $0.5$, while configurations containing age fall to $0.531$--$0.679$	(Section~\ref{sec:res-matched}). 
	
	None of the sixteen paired comparisons computed on the full
	cohort excludes zero, and every configuration---including a one-predictor age
	model---clears the conventional chance threshold. The conclusion we draw is
	therefore about evaluation before it is about voice: a non-age acoustic signal
	is present and measurable in these data, and the protocol in standard use cannot
	see it.
	
	We do not claim that voice-based COPD screening is clinically viable. The
	matched cohorts contain roughly two dozen pairs drawn from a single site, a
	single language, and a single elicitation task, and the conformal analysis of
	Section~\ref{sec:res-aux} indicates that nominal $90\,\%$ coverage requires
	declining to commit on most participants.
	
	\subsection{Implications for Cross-Sectional Screening Studies}
	
	The failure mode we document is not specific to this corpus. It requires only
	two conditions: that cases and controls differ demographically, and that
	descriptive statistics be computed over recordings rather than over
	participants. Both are common. Any repeated-measures speech dataset with unequal
	per-participant contributions admits the inversion of Fig.~\ref{fig:age}(c), and the direction of the distortion cannot be anticipated---here a single control contributing a quarter of the corpus reversed the sign of the group age difference.
	
	Two useful practices follow. First, cohort balance should be reported at the unit of
	analysis used by the classifier. Where a model issues one prediction per
	participant, participant-level balance is the relevant quantity, and
	recording-level summaries may conceal the imbalance they are intended to
	characterise. Second, and more consequential, the confounder's own
	discrimination should be reported on the cohort used for evaluation. A balance
	statistic such as Eq.~\eqref{eq:smd} compares means, whereas the metrics used to
	assess classifiers are rank statistics. Thus, the two can disagree.
	
	This bears on how the literature surveyed in Section~\ref{sec:rw-cross} should
	be read. Reported cross-sectional accuracies in the $75$--$95\,\%$ range are not
	in question as measurements: what is unclear is their composition. Where cases
	are older than controls and age is either supplied to the model or encoded in
	age-correlated acoustics, the reported figure is an upper bound whose split
	between pathology and demographics is unknown. The within-patient longitudinal
	designs~\cite{vanbemmel2021features,nallanthighal2022exacerbation,vanbemmel2026tacticas,mayr2024clinical,bhalla2025wearable}	are largely immune to this, which is consistent with their reporting more physiologically grounded and more modest effects. The two literatures should not be cited as joint evidence for a single proposition.
	
	\subsection{Confounded Covariates Exclusion}
	
	A potentially transferable finding is that supplying a confounded covariate does not
	leave the rest of the model intact. Among age-containing configurations, the performance lost to matching scales inversely with the acoustic information available alongside the same demographic block. Along the \texttt{DAC} ladder the drop falls monotonically
	from $-0.178$ at one predictor to $-0.121$ at two, $-0.086$ at $51$ and $-0.062$
	at $103$; along the \texttt{ALL} ladder from $-0.178$ to $-0.129$ at six, $-0.097$ at $55$ and $-0.085$ at $107$. Permutation importance measures the same thing directly: in \texttt{ALL}, age ranks first with more than twice the importance of the leading acoustic descriptor.
	
	The mechanism is unsurprising in retrospect. A gradient-boosted ensemble
	allocates capacity to whichever predictors reduce training loss fastest, and on
	a cohort where age separates the classes it will exploit age in preference to a
	weaker, noisier acoustic signal. The acoustic representation is then
	under-learned, and the deficit is exposed as soon as the confound is
	neutralised. Consequently, reporting a no-age ablation of a model trained
	\emph{with} age---the standard defence---understates the acoustic signal, because
	the ablated model never had to learn it. The configurations trained without age outperform their age-containing counterparts on matched cohorts by $0.056$ and $0.099$ ROC-AUC.
	
	This argues against the common practice of including demographics as routine
	covariates in screening models. Where a demographic variable is a known
	confounder of the recruitment design rather than a causal contributor, excluding
	it at training time is not merely a robustness check but a modelling decision
	with measurable consequences.
	
	\subsection{Recording Conditions}
	
	With age matched and verified, gender measured at $0.479$ on the matched
	cohorts, phonation duration removable without altering the conclusion, and the
	symptom-questionnaire block contributing nothing detectable, the leading
	non-pathological explanation for the residual signal is confounding by
	recording conditions: device model, microphone, acoustic environment, gain,
	session, or operator.
	
	We are unable to exclude it, and we regard this as the study's most substantive
	limitation. Such confounding would produce precisely the signature we observe.
	It is orthogonal to participant age, so it would not get affected by age matching. Also, it is invisible to demographic controls and it would express itself in spectral
	and perturbation descriptors, which is where we locate the signal. Recruitment
	via clinical pathways for cases and other channels for controls can introduce
	systematic differences in device or setting without any deliberate design
	choice. The released feature table does not permit a test. Only two indirect checks are
	available and both are reassuring but weak: recording count does not
	discriminate (ROC-AUC $0.545$), indicating no gross difference in study
	adherence between groups, and removing phonation duration leaves the matched
	result intact. Neither addresses channel effects.
	
	Thus, we recommend that corpora intended as cross-sectional benchmarks record and release all metadata as a matter of course. Applied to the present result, the appropriate reading is that a non-age, non-gender acoustic signal is present, and that whether it is respiratory or instrumental in origin remains open.
	
	\subsection{Relation to Speaker Disentanglement}
	
	The approach closest to ours in intent is adversarial suppression of speaker
	identity in the learned representation~\cite{yan2026disentangled}, which
	improves diagnostic accuracy while reducing measured identity leakage. That work
	and this one address different levels of the same problem and compose rather
	than compete. Disentanglement operates on the individual: it prevents a model
	from recognising \emph{which} speaker is talking. Matched evaluation operates on
	the group: it prevents a model from being credited for discriminating on a
	covariate that differs between cohorts. A representation from which individual
	identity has been removed may still encode age, and where cases are
	systematically older than controls, age remains predictive of the label. The
	converse also holds---matched evaluation does not prevent identity leakage
	within a longitudinal design. A study exposed to both threats needs both
	remedies, and both need to be measured rather than assumed.
	
	\subsection{Limitations}
	
	Beyond recording-condition confounding, which we consider the principal
	limitation, the following constrain the conclusions.
	
	The matched cohorts retain $24.0$ pairs on average, $48$ of $67$ participants,
	and the discarded participants are systematically the younger controls. Matched and pooled estimates are therefore different estimands rather than
	corrected versions of one another. Because the bootstrap resamples matched pairs
	rather than individuals, the effective number of independent units is $24$, and
	the intervals are correspondingly wide---approximately $0.33$ for the age-free
	configurations, against a Hanley--McNeil~\cite{hanley1982meaning} expectation of $0.30$ at $24$ versus $24$ independent observations. Two of the five age-free configurations
	consequently fail to exclude chance on ROC-AUC despite point estimates near
	$0.68$.
	
	Matching neutralizes age to a measurable but not perfect degree. 
	Thus, a	small non-monotone residual is still present (but not measured in this work) despite the two-year caliper. Research towards measuring this residual would be an interesting direction of work. 
	
	Matching is applied at evaluation time to models trained on the unmatched
	cohort. This tests whether the learned signal is present after the removal of age
	information but it does not establish what a model trained under balance would
	learn. Age-stratified or matched training is a natural extension.
	
	Twelve configurations were evaluated without multiplicity correction. Only one paired contrast against raw age excludes zero ($+0.207$ $[+0.027, +0.365]$). The claims in Section~\ref{sec:res-matched} rest on against-chance intervals, which do not
	depend on that contrast. 
	
	Hyperparameter selection by the one-standard-error rule resolved to the
	minimum-capacity grid point in nearly every split, because inner-fold AUC is
	close to flat across the grid at this sample size. Model capacity was therefore
	effectively fixed rather than tuned. We regard this as benign: nested
	cross-validation estimates whatever procedure is specified, all configurations
	were treated identically, and the replication with fixed hyperparameters agrees on the family contrast	(Pearson $r = 0.902$ across configurations). At the same time, it means the reported estimates should not be read as the best achievable with this model family.
	
	Finally, the cohort is single-site, single-language, and restricted to sustained
	phonation of one vowel, with $30$ cases. Generalization across languages,
	elicitation tasks, recording channels and disease severity is untested, and the
	conformal analysis indicates the models are far from clinically decisive.
	
	\section{Conclusions}
	\label{sec:conclusions}
	
	We re-evaluated a public sustained-phonation corpus for COPD classification
	under a strictly participant-level protocol, and found that the cohort, although
	described as matched on age within five years, exhibits a participant-level age
	standardized mean difference of $0.726$. Also, eight of $37$ controls have no case
	within five years, and published figures are consistent with recording-level rather than participant-level weighting, which inverts the sign of the group age difference.
	
	Under this imbalance, conventional pooled evaluation is uninformative. All
	twelve feature configurations we examined clear the chance threshold, a
	hand-specified non-monotone function of age alone attains ROC-AUC $0.721$, and
	none of sixteen paired comparisons excludes zero.
	
	Evaluating instead on repeatedly drawn age-matched cohorts, and reporting the
	discrimination achieved by raw age and raw gender on those same cohorts,
	separates the two model families. Where both confounders are measured at chance,
	acoustic models excluding age retain ROC-AUC $0.717$ $[0.552, 0.859]$ and
	average precision $0.747$ $[0.581, 0.892]$, while models containing age fall
	towards chance. 
	
	Two findings have broader methodological implications. Supplying a confounded covariate at training time measurably reduces transfer to age-balanced evaluation cohorts, so
	excluding it is a modeling decision rather than a robustness check. Moreover, a
	compact set of fourteen classical voice-quality and perturbation measures achieves comparable discrimination to a $55$-dimensional combined representation, with better-behaved importance estimates and greater decisiveness under conformal prediction.
	
	We conclude that a non-age, non-gender acoustic signal is present and measurable
	in these recordings, that its origin, respiratory or instrumental, cannot be
	settled without recording-condition metadata that the corpus does not release,
	and that the evaluation protocol in standard use for cross-sectional voice
	screening cannot distinguish these possibilities. Reporting participant-level
	cohort balance, and the confounder's own discrimination on the evaluation
	cohort, would be inexpensive and would materially improve the interpretability
	of this literature.
	
	\section*{Acknowledgments}
	The authors would like to sincerely thank A. Idrisoglu, A. L. Dallora, A. Cheddad, P. Anderberg, A. Jakobsson, and J. S. Berglund for sharing the COPDVD dataset in the form of pre-computed features.
	
	\ifdefined\HASIEEE\bibliographystyle{IEEEtran}\else\bibliographystyle{plain}\fi
	\bibliography{references}
	
\end{document}